\documentclass{sig-alternate-05-2015}

\usepackage[labelformat=simple]{subcaption}
\usepackage{tikz}
\usepackage{array}
\usepackage{multirow}
\usepackage{tabularx,colortbl}

\let\oldnl\nl
\newcommand{\nonl}{\renewcommand{\nl}{\let\nl\oldnl}}

\newtheorem{definition}{Definition}
\newtheorem{example}{Example}

\begin{document}

\setcopyright{acmcopyright}

\doi{10.475/123_4}

\isbn{123-4567-24-567/08/06}


\acmPrice{\$15.00}

%

\title{Leveraging Neighborhood Summaries for Efficient Queries on Existing Relational RDF Systems}

\numberofauthors{3} 
\author{
      \alignauthor Lei Gai\\      
      \email{lei.gai@pku.edu.cn}
\end{tabular}\newline\begin{tabular}{c}
      \affaddr{School of Electrical Engineering and Computer Science}  \\
      \affaddr{Peking University. Beijing, P.R.China}   \\
}

\maketitle

%
%

\begin{abstract}
Using structural informations to summarize graph-structured RDF data is helpful in tackling query performance issues. However, leveraging structural indexes needs to revise or even redesign the internal of RDF systems. Given an RDF dataset that have already been bulk loaded into a relational RDF system, we aim at improving the query performance on such systems. We do so by summarizing neighborhood structures and encoding them into triples which can be managed along side the exist instance data. At query time, we optimally select the effective structural patterns, and adding these patterns to the existing queries to gain an improved query performance. Empirical evaluations shown the effectiveness of our method. 

\end{abstract}


%
%
\printccsdesc


%
%

\section{Introduction}

The Resource Description Framework(RDF) have been pervasively adopted in many fields to represent the extensively linked resources. The available RDF data keep increasing in both size and quality, this challenges for efficient query processing of such graph-structured data.  Although join-based query processing over relational RDF stores can benefit greatly from 40+ years of advances in database domain, the problem of losing graph structure in relational RDF representation is still not well resolved. It is well established that the summaries of structural informations is useful in pruning irrelevant triples, and therefore can greatly improve RDF query performance.  Existing researches on structural indexes for RDF data are mainly focused on building specialized index structures (e.g., \cite{DBLP:conf/www/PhamPEB15,DBLP:journals/tkde/TranLR13}), or even need to build query-specific system from scratch (e.g., \cite{Zou:2014:GGS:2647883.2647907}). However, in most cases the bulk of dataset have already been managed by an existing RDF stores, reload or make another copy of data is a waste of resources. There is a need to improve the query performance in an existing system without further revision of its internals.   

In this paper, we alleviate this problem by summarizing the adjacency graph structures as triples, and introduce a novel mechanism that leverage neighborhood summaries in rewriting equivalent queries. Such queries are expected to be more efficient when executed on existing query engines. The contributions of our work are mainly two-fold:

First,  we propose a method that encode the structural informations as triples. This design entails seamless integrate and index of graph structures. Generally, data hold in relational backends are indexed by values. Consider that graph-shaped queries are more selective by structures than by values, index data by structures can improve query performance to a certain extends.We avoid modifying the internal of RDF stores by encoding structural informations as triples. For the structural summarization to be cost-effective in both construction and storage, we adopt the neighborhood structures which can be built in linear time. The scale of the data can be further refined by applying structural merging strategies. 

Second, we identify neighborhood summaries as patterns and optimally reform queries at runtime. Minimization of intermediate results is a prominent problem in join-based query optimization. The introduction of such patterns in queries are useful in pruning irrelevant intermediate results, thus  can potentially gain query performance improvement. For sake of optimal selection of such patterns, we resort to maintain a carefully cultivated in-memory metadata to check the conditions. The reformed queries  can be executed on all existing SPARQL-compliant query engines.

To summarize, our method offers the following features: 1) Schema independent integration of RDF structural informations, 2) efficient query processing by leveraging cost-effective neighborhood structural summaries.  We further evaluated our method on benchmark and DBpedia datasets. Results shown the correctness and efficiency of our method. 

The rest of the paper is organized as follows:  \textit{Section} \ref{sec-basic} gives an overview of our method, \textit{Section} \ref{sec:offline} introduces how to maintains structural summaries, \textit{Section} \ref{sec:naqr} explains the consideration of optimal query reformulation, \textit{Section} \ref{sec:experiments} presents the preliminary evaluation results, and \textit{Section} \ref{sec:conclusions} concluded.

%
%

\textbf{Related Works: }\label{sec:related} Structural index is a kind of graph summarization that was first introduced by \cite{DBLP:conf/icdt/MiloS99} almost 20 years ago. Many researches have been devoted to semi-structured XML data and more recently to graph-structured RDF data. According to the usage of structural indexes, we categorize existing researches that are most related to our works into three paradigms: 1) Queries are fully performed on structural indexes, e.g., \cite{Zou:2014:GGS:2647883.2647907}, 2) using structural indexes for filtering, then perform query on the reduced data, e.g., \cite{DBLP:journals/tkde/TranLR13}, and 3) queries is partially performed on indexes for structural manageable data, e.g., \cite{DBLP:conf/www/PhamPEB15}. Our work follows the third paradigm, differ in that we consider summarizing all the neighborhood structures instead of bi-simulation-based summarization, and store them along side the original data. This entails an cost-effective structural indexes. Furthermore, all existing researches are not schema-agnostic, this means they need to do extra works on modifying existing systems. Our method is more pragmatic in that it can be applied to any existing relational RDF systems.

%
%

\section{Basic Concepts and Overview}\label{sec-basic}

In an RDF dataset $\mathcal{D}= \bigcup^{|\mathcal{D}|}_{i=1} t_i$, each triple $t_i$ is uniformly represented as three elements $\langle$\textit{Subject, Predicate, Object}$\rangle$, or $\langle$s,p,o$\rangle$ for brevity. $\mathcal{D}$ can be naturally modeled as a directed, edge-labeled \textit{RDF graph} $\mathcal{G}=\langle V,E \rangle$. Each triple $t_i \in \mathcal{D}$ corresponds to a directed labeled edge $e \in E$. Adopting query language such as SPARQL\footnote{\scriptsize{\url{https://www.w3.org/TR/sparql11-query/}}}, a conjunctive RDF query $Q$ can be expressed as: \textbf{SELECT} $?v_1,\ldots$ \textbf{FROM} $TP$, where $TP=\{tp_1,\ldots, tp_k\}$ is a combination of \textit{triple pattern}s, each one of which has a least one of \textit{s}, \textit{o} or \textit{p} replaced by variables $?v$. 

\vspace*{-5pt}
\begin{definition}[Neighborhood Summary]
A \textit{Neighborhood Summary (NS)} in $\mathcal{D}$ is the set of distinct \textit{Predicte}s in triples which share a common \textit{Subject}/\textit{Object} $v$. Formally, $NS(v)=\{p| t=\langle s/v,p,o/v\rangle, \forall t \in \mathcal{D}, i \neq j \rightarrow p_i \neq p_j \}$, and the element $v$ is called a \textit{pivot}. Similarly, a \textit{Neighborhood Summary} $NS_q(v_q)$ in $Q$  is the set of distinct \textit{Predicte}s in \textit{triple pattern}s which share a common \textit{Subject}/\textit{Object} $v_q$.
\end{definition}

From graph perspective, $NS(v)$ depicts all the distinct label of the the incoming and outgoing edges of vertex $v$ in $\mathcal{G}$. In this paper, we encode each $NS(v)$ in $\mathcal{D}$ with a distinct URI $l_{NS}(v)$, and designate a specialized \textit{Prediate} "ssf:pivot". 

\vspace*{-5pt}
\begin{definition}[NS-Triple, NS-Pattern]
A \textit{NS-Triple} is the triple in forms of $t(v)=\langle v,"ssf:pivot",l_{NS}(v) \rangle$. A \textit{NS-Pattern} of $NS_q(v_q)$ for $Q$ is the triple pattern $tp(v_q)=\langle v,"ssf:pivot",l_{NS_q}(v_q) \rangle$.
\end{definition}

\begin{figure}{}
  \centering
  \includegraphics[height=0.20\textheight]{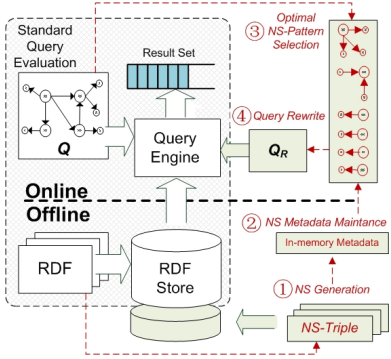}
\caption{System overview.}
\label{fig:overview}
\end{figure}
\vspace*{-5pt}

An Overview of our method is depicted in \textit{Figure} \ref{fig:overview}. Generally, relational query processing follows a \textit{Scan-Join} paradigm. $\mathcal{D}$ can be viewed as managed in a single large SPO table $\mathcal{T}$ in RDF stores. The candidate set of each triple pattern $tp_i \in Q$ is generated by a scan of $\mathcal{T}$. The query engine determines an optimal join order for all candidate sets, aiming at minimize the intermediate results size. The final query results is derived by the actual execution of joins. In our method, we pre-process $\mathcal{D}$ to generate all NSs, and encode them as a set of NS-Triples (step \textcircled{1}, detailed in \textit{Section} \ref{sec:offline}). These NS-Triples are managed by the RDF store where $\mathcal{D}$ are stored. Meanwhile, we introduce a dedicated designed in-memory structure to maintain the statistics (i.e. metadata) of NS-Triples (step \textcircled{2}, detailed in \textit{Section} \ref{sec:naqr}). Henceforth, once an ad hoc query $Q$ is committed at runtime, we detect and determine the optimal NS-Patterns to be used, and reform an equivalent query $Q_R$ by adding these NS-Patterns (step \textcircled{3}, detailed in \textit{Section} \ref{sec:naqr}). Consecutively, $Q_R$ is processed by the query engine and return the final results (step \textcircled{4}). As $Q_R$ considers the structural characteristics in query, as well as the trade-off of the extra cost introduced by joining NS-patterns, it is expected to  gain an enhanced query performance.

%
%

\section{Refined Neighborhood Summary} \label{sec:offline}

We present in this section an approach of using NS-Triples to shred the adjacency structure summaries into RDF triples. NS-Triples can be stored alongside the corresponding instance data $\mathcal{D}$. Consequently, existing SPARQL query engines can take advantage of summary-based optimizations without modifying the internals of SPARQL engines. The generation of NS-Triples follows three steps:

1) \textbf{Basic NS generation}. In this step, the occurrence of all NSs are detected. As stated before, $\mathcal{D}$ have been bulk loaded into an existing RDF store, and can be viewed as a single SPO table $\mathcal{T}$. Ideally, we make a first scan through $\mathcal{T}$ following the order of \textit{Subject}, and maintain a hash map $H$ that take each \textit{pivot} as key, while the value is the set of distinct \textit{Predicts}s in triples that has \textit{pivot} as \textit{Subject}. Afterwards, we make a second scan through $\mathcal{T}$ following the order of \textit{Object}. For each non-literal \textit{Object}, If it is already exists as key in $H$, append the associated distinct \textit{Predicate} to its value set. Otherwise, add an new entry of $H$ with \textit{Object} as the key and append \textit{Predicate}s into value set accordingly. In this way, we get all the \textit{pivot}s and its NS in $H$. To get the metadata $\mathcal{M}_H$ that holds all distinct NS and its frequency of occurrence, a scan of $H$ is needed that sort distinct NS and count the number of \textit{pivot} that has the same NS. Obviously, the scan of $T$ is linear in average-case complexity. The cost of this procedure is mainly up to the construction of $\mathcal{M}_H$, which is $O(nlogn)$ where n is the number of \textit{pivot}s.
 
2) \textbf{NS refinement}. Theoretically, there may exists a great number of distinct NSs considers the number of $|P|$  can be rather large (e.g., currently DBpeida has 60,223 different \textit{Predicate}s). Therefore, maintaining and retrieve of $\mathcal{M}_H$ can be a great cost. We adopt a simply strategy to structurally merge NSs. Let $\{NS\}$ denotes the set of \textit{pivot} that has NS, if $NS_i \subset NS_j$, and $\{NS_i\}=\{NS_j\}$, then $NS_i$ can be merged into $NS_j$. In practice, this strategy can greatly reduce the size of $\mathcal{M}_H$.

3) \textbf{NS-Triple generation}. To encode a NS in a triple, first we represent the set of $NS(v)$ by an URI $l_{NS}(v)$, which is generated by a hash function. The generated $l_{NS}(v)$ are also kept in $\mathcal{M}_H$. Then we use a dedicated \textit{Predicate} "ssf:pivot"  and generate \textit{NS-Triple} in forms of $t(v)=\langle v,"ssf:pivot",l_{NS}(v) \rangle$.

\begin{figure*}
\captionsetup{belowskip=0pt,aboveskip=0pt}
\captionsetup[sub]{margin=0.1pt,skip=0.1pt, labelfont=bf}
\centering
\begin{subfigure}{0.64\textwidth}
  \includegraphics[height=0.09\textheight,width=0.9\textwidth]{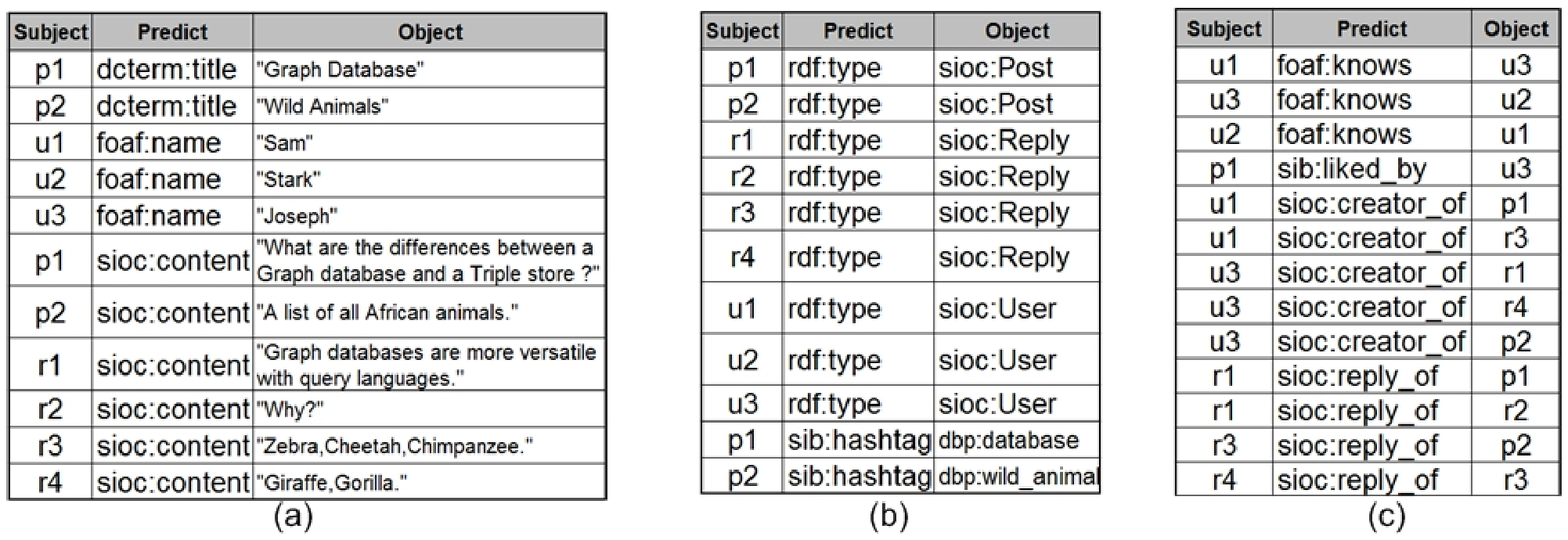}
  \subcaption{List of triples in RDF dataset $\mathcal{D}$.}
  \label{subfig-triplelist}
  \par\medskip 
  \includegraphics[height=0.16\textheight,width=0.9\textwidth]{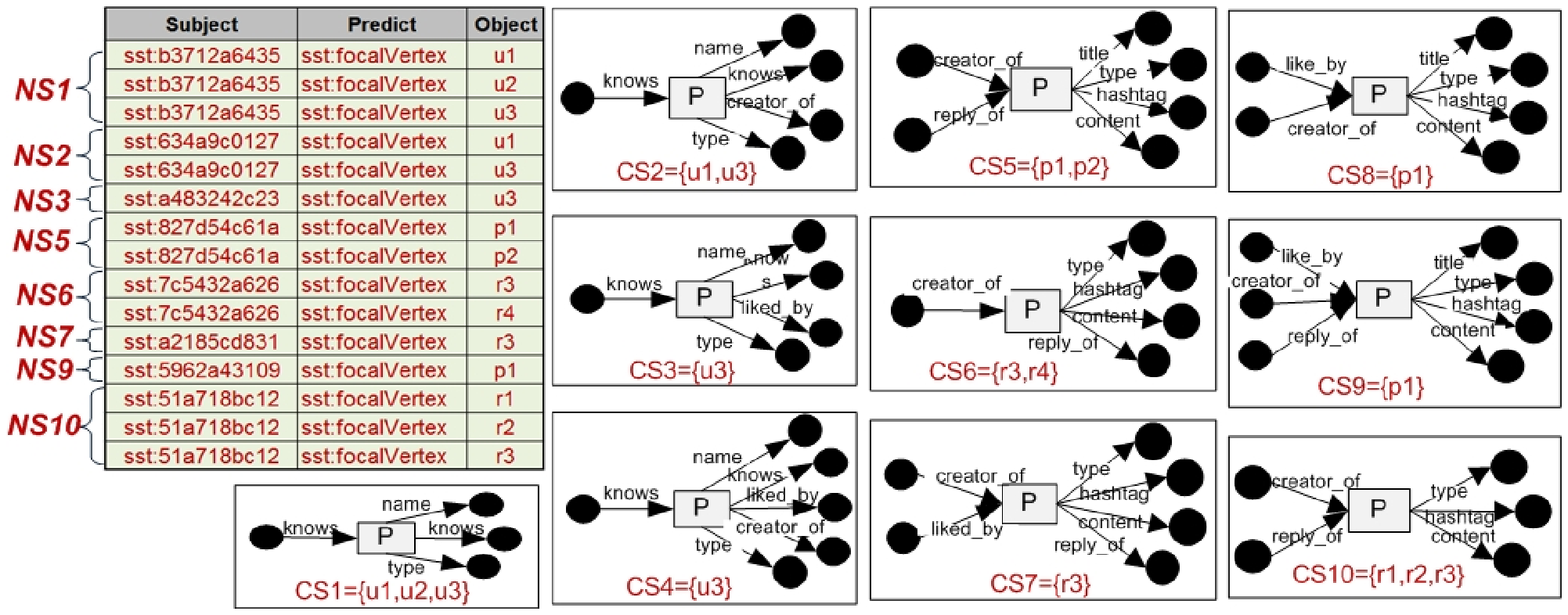}
  \subcaption{Neighborhood Summaries (NS) for $\mathcal{D}$, and list of NS-Triples.}
  \label{subfig-cs}
\end{subfigure}
\hspace*{\fill}
\begin{subfigure}{0.30\textwidth}
  \vspace*{\fill}
  \includegraphics[height=0.25\textheight,width=\textwidth]{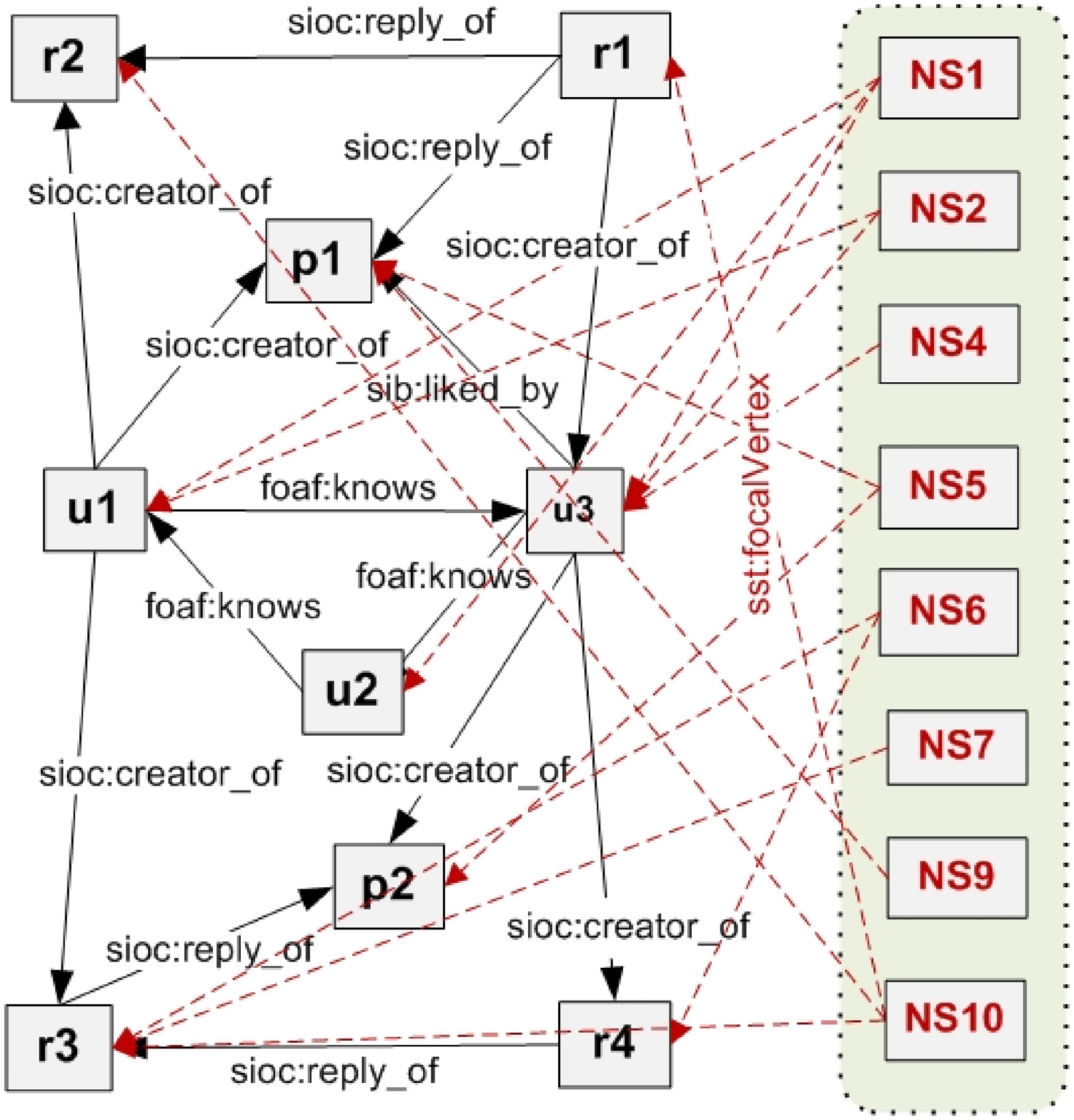}
  \subcaption{ $\mathcal{G}$, the RDF graph of $\mathcal{D}$.}
  \label{subfig:rdf-graph}
\end{subfigure}
\hspace*{\fill}

\caption{An exemplar RDF dataset $\mathcal{D}$, its data graph $\mathcal{G}$ and Character Sets }
\label{fig:example}
\end{figure*}

\vspace*{-5pt}
\begin{example}
\begin{scriptsize}
\textit{Figure} \ref{subfig-triplelist} shows the list of triples in an example twitter-like dataset $\mathcal{D}$. By scanning $\mathcal{D}$, we can get 10 distinct NSs as shown in \textit{Figure} \ref{subfig-cs}. Using the defined NS refinement strategy, we can merge NS3 into NS4, and NS8 into NS9. The generated NS-Triples are also listed in \textit{Figure} \ref{subfig-cs}. \textit{Figure} \ref{subfig:rdf-graph} shows the graph representation of all \textit{pivot}s and NS-Triples. 
\end{scriptsize}
\end{example}

%
%

\section{Neighborhood-aware Query Reformulation} \label{sec:naqr}

In essence, NS-Triples capture correlations between triples that share common \textit{pivot}. A native strategy is to simply identifies all the neighborhood structures $NS_Q$ in $Q$ to generate NS-Patterns, adds all NS-Patterns accordingly. This entails us to directly add NS-Patterns in $Q$ to "pruning" the sets of candidates. However, it may not always works well. Two problems arises that hinder the efficiency of our method.

The first problem is in finding the minimal NS in $\mathcal{M}_H$ that contains a specific NS in $Q$. This serves as determining the \textit{Object} value of the generated NS-Pattern.  We name this procedure as \textit{NSMatch()}. Formally, \textit{NSMatch()} find $NS_i^m$ in $\mathcal{M}_H$, where $NS_i^m \supseteq NS_i^Q$, and if $\exists NS_j \in NS,  NS_j \supseteq NS_i^Q$, then $NS_j \supset NS_i^m$.  Recall that in the offline processing stage, we have maintained an in-memory metadata $\mathcal{M}_H$that contains $\forall NS_i \in NS$, its identifier $l_{NS}(i)$ and frequency $f(i)$. Consider that $\mathcal{M}_H$ may be rather large, we need to organize the structure of $\mathcal{M}_H$ for efficient retrieval of set-containment-based NS matching. Basically, we observe that the content of NS set represents a tree shape, thereby we can organize $\mathcal{M}_H$ in a set of tree structures to facilitate \textit{NSMatch()} retrievals. To elaborate, the root of a tree is the common \textit{Predicate}s that a set of $NS_i \in NS$ shares, and each $NS_i$ is placed on the tree leaf at different level following the increment of NS set content. Consequently, \textit{NSMatch()} can be reduced to find the tree by matching the root, then matching along side this tree until no further leafs can be matched. Thereby the current leaf node is the minimal $NS_i^m$ that contains $NS_i^Q$. Having $NS_i^m$, we can get $l_{NS}(i)$ from $\mathcal{M}_H$, and generate the NS-Pattern $\langle l_{NS}(i),"ssf:pivotVertex",v_q \rangle$.

Second, note that the main idea behind our approach is to add neighborhood structures as patterns in a query to "pruning" the sets of candidates. This strategy is effective when such "pruning" outweighs the cost of NS-Pattern processing. Intuitively,  the naive strategy may fail due to the following cases:

\vspace*{-5pt}
\begin{itemize}
\item $Q$ has high-selective triple patterns. As such high-selective TPs can essentially reduce the candidate sets, there might be no need to add extra NS-Patterns.
\item $Q$ has relatively low-selective NS-Patterns. In this case, the cost of join operation introduced by adding extra NS-Patterns can easily exceed that of the original triple pattern joins.
\end{itemize}

To avoid the above drawbacks, we need to introduce additional statistics in metadata. A native idea is to hold all statistics of $\mathcal{D}$ as well as NS-Patterns in memory, this may cause great memory consumptions consider the sheer size of $\mathcal{D}$. In reality, we can roughly estimate whether the candidate NS-Patterns to be added has a relatively high selectivity than all its related triple patterns in $Q$. In view of that, we only maintain the non-literal elements which has selectivity greater than a given threshold $\delta$. Given that we can roughly estimate the selectivity of a triple pattern following the independence assumption at runtime. If the elements in a pattern is not held in such metadata, or its estimated selectivity is relatively lower than that of its related NS-Pattern, this means that adding this NS-Pattern potentially introduces high cost than its pruning ability. Then there is no need to add this NS-Pattern.

\vspace*{-5pt}
\begin{example}
\begin{scriptsize}
Given a query $Q=$"SELECT ?x ?y ?z WHERE \{ $\overbrace{?x\ creator\_of\ ?y}^{TP1}$. $\overbrace{?y\ type\ Reply}^{TP2}$. $\overbrace{?y\ reply\_of\ ?z.}^{TP3}$\}", clearly TP1-3 forms a NS-Pattern $NS(?y)$. Using \textit{NSMatch()}, we can get NS6 instead of NS7 as the minimal match of $NS(?y)$. By maintaining the necessary statistics, we can roughly estimate the selectivity as $Sel(TP1)=9/50$, $Sel(TP2)=4/50$ and $Sel(TP3)=4/50$, while $Sel(NS(?y))=2/50$. As NS(?y) gives the relative higher selectivity, we can optimally rewrite $Q$ by adding NS-Pattern $\langle ?y
,"nss:pivotVertex", "nss:7c5432a626" \rangle$.
\end{scriptsize}
\end{example}

%
%

\section{Preliminary Results}\label{sec:experiments}

\begin{figure*}
\captionsetup{belowskip=0pt,aboveskip=0pt}
\centering
  \includegraphics[height=0.14\textheight,width=\textwidth]{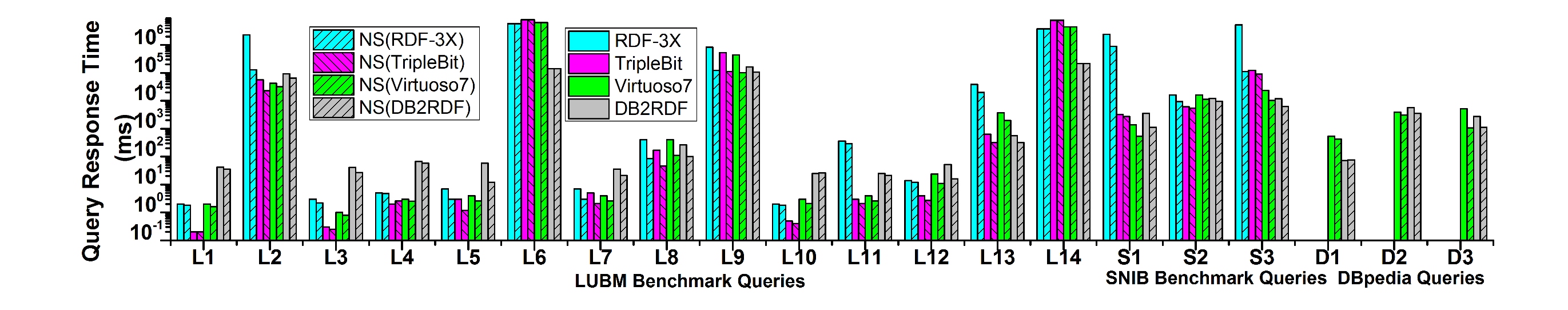}
\caption{Query response times on existing systems. }
\label{fig:evaluation}
\end{figure*}

Our evaluations focused on the quantitative metrics, i.e. the query response time. All experiments were performed on a machine with Debian 7.4 in 64-bit Linux kernel, two Intel Xeon E5-2640 2.0GHz processors and 64GB RAM. We empirically evaluated the RDF query performance on two kinds of large-scale benchmark data, LUBM\footnote{\scriptsize{\url{http://swat.cse.lehigh.edu/projects/lubm/}}} generated with 20,480 universities, SNIB\footnote{\scriptsize{\url{http://www.w3.org/wiki/Social_Network_Intelligence_BenchMark}}} of 15,000 users, and real DBpedia dataset\footnote{\scriptsize{Downloaded from \url{http://wiki.dbpedia.org/Downloads2015-04}}}. The statistics are shown in \textit{Table} \ref{tbl:datastats},  along with the increased percentage of triples after applying NS-Triple generation. Results shown a relative small extra storage costs. The amount of generated triples are mainly up to the structural complexity of data (e.g. Twitter-alike SNIB dataset) and the number of distinct \textit{Predicate}s (e.g. DBpedia).

\vspace*{-5pt}
\begin{table}[ht]
\captionsetup{belowskip=0pt,aboveskip=0pt}
\caption{Statistics and characteristics of RDF dataset.}
\label{tbl:datastats}
\scriptsize
\centering
    \begin{tabular}{ |>{\raggedright\arraybackslash}m{0.75in} | >{\centering\arraybackslash}m{0.6in} |  >{\centering\arraybackslash}m{0.6in} | >{\centering\arraybackslash}m{0.6in} |}
    \hline
    Dataset   &   \textbf{LUBM}   &   \textbf{SNIB}   & \textbf{DBpedia} \\
    \hline
    \# of Triples      &   4,845M   &   1,774M  & 489M   \\
    \hline
    \# of Predicates   &     18   &  48  & 60,223 \\
    \hline
    \# of NSs   &     26   &  47  & 172,927 \\
    \hline
    increase in size   &     14\%   &  19\%  & 37\% \\
    \hline
   \end{tabular}
\end{table}

We chosen four representative systems as existing RDF stores. Among them, RDF-3X\footnote{\scriptsize{Available at \url{https://github.com/gh-rdf3x/gh-rdf3x/}}}\cite{rdf3x2010} and  T{}ripleBit\footnote{\scriptsize{Available at \url{http://grid.hust.edu.cn/triplebit/TripleBit.tar.gz}}}\cite{triplebit-vldb2013} are dedicated systems which indexed all possible permutation of s,p and o. Virtuoso 7\footnote{\scriptsize{Available at \url{https://github.com/openlink/virtuoso-opensource}}} and DB2RDF\footnote{\scriptsize{Available at \url{https://github.com/Quetzal-RDF/quetzal/}}}\cite{DBLP:conf/sigmod/BorneaDKSDUB13} are RDB-based system which use commercial-off-the-shelf RDBMS as backends. Each query was executed 11 times in a consecutive manner in each system, and the first time of each round was used for cache warm-up. We reported the arithmetic mean of the rest 10 queries in \textit{Figure} \ref{fig:evaluation}. Results shown the performance enhancement for most queries (ranging from 12\% for L3, L4, to 213\% for S3). The effect of our method is especially obvious for more complex queries that contain relatively more triple patterns (e.g. L2, L7, L8, L9, L12 for LUBM, and all SNIB queries). Among the four systems, RDF-3X gained the most prominent improvement in query performance, this can be reduced to that its query engine which use selectivity-based optimization could benefit greatly from the reduction of intermediate results brought by NS-Patterns. Note that our method had no effects on some queries (e.g., L1, L6, L14, D1), but also did not downgrade theirs original performance. This shown the effectiveness of optimal NS-Patterns selection.  


%
%
\vspace{-5pt}
\section{Conclusions}\label{sec:conclusions}

In this paper, we introduced the notion of \textit{Neighborhood Summary} for optimized query processing on an existing relational RDF systems. We intended to use this cost-effective structural information in helping prune the irrelevant intermediate results to boost query performance. Preliminary experiments shown the effectiveness of our method.

\textbf{Acknowledgments.} This work is supported by <TBD>.

%
%

\bibliographystyle{abbrv}
\bibliography{edbt-2017}  

%
%
%
%

\vspace*{-5pt}
\section{Appendix}\label{sec-appendix}


\noindent \textbf{SNIB Queries} \label{appendix-snibquery}
\begin{scriptsize}








\vspace{0em} \noindent \textbf{S1:}SELECT ?user ?commentcontent ?commentdate WHERE\{?user foaf:knows ?friend.
 ?user rdf:type sib:User.
 ?friend rdf:type sib:User.
 ?friend sioc:moderator\_of ?forum.
 ?post sib:hashtag dbp:Creek .
 ?post rdf:type sioc:Post.
 ?forum sioc:container\_of ?post.
 ?post sioc:content ?postcontent.
 ?post sioc:container\_of ?postcomment.
 ?postcomment rdf:type sioc:Item.
 ?postcomment sioc:content ?commentcontent.
 ?postcomment dc:created ?commentdate. \}

\vspace{0em} \noindent \textbf{S2:}SELECT ?user1 ?user2 ?friend WHERE\{?user1 foaf:knows ?user2.
 ?user2 foaf:knows ?friend .
 ?friend foaf:knows ?user1 .
 ?user1 rdf:type sib:User .
 ?user2 rdf:type sib:User .
 ?friend rdf:type sib:User .
 ?friend sioc:creator\_of ?post.
 ?post rdf:type sioc:Post .
 ?post sioc:content ?postcontent.
 ?post sib:hashtag dbp:Creek .
 ?post sib:liked\_by ?user1\}

\vspace{0em} \noindent \textbf{S5:}SELECT ?u1 ?u2 WHERE\{?f sib:memb ?u1.
 ?f sib:memb ?u2.
 ?u2 foaf:knows ?u1.
 ?u3 foaf:knows ?u2.
 ?u3 foaf:knows ?u1.
 ?u1 rdf:type sib:User.
 ?u2 rdf:type sib:User.
 ?u3 rdf:type sib:User.
 ?p sib:hashtag dbp:Hummelshof.
 ?u1 sioc:creator\_of ?p.
 ?p rdf:type sioc:Post.
 ?forum sioc:container\_of ?p.
 ?forum sioc:container\_of ?c.
 ?c rdf:type sioc:Item.
 ?u2 sioc:creator\_of ?c.
 ?c sioc:reply\_of ?p.
 ?p sib:liked\_by ?u2. \}

\end{scriptsize}


\noindent \textbf{DBpedia Queries} \label{appendix-snibquery}
\begin{scriptsize}








\vspace{0em} \noindent \textbf{D1:}SELECT ?club WHERE \{ 
     ?club  p:league r:Premier\_League. 
     ?club p:league  r:Scottish\_Premier\_League.  
     ?club p:league  r:Football \_League\_One. 
     ?club p:league  r:Football \_League \_Championship. 
     ?club p:league  r:Football \_League \_Two.
  \}

\vspace{0em} \noindent \textbf{D2:}SELECT ?v1 ?v2 WHERE \{ 
\{ ?v1 rdf:type o:Settlement.
?v1 p:population ?v2. 
FILTER ( xsd:integer(?v2) > 54 ) \} 
UNION \{ 
?v1 rdf:type o:Settlement.
?v1 p:populationUrban ?v2.
 FILTER ( xsd:integer(?v2) > 54 ) 
\} \}

\vspace{0em} \noindent \textbf{D3:}SELECT ?V1 ?v2 ?v3 WHERE \{  
?v1 rdf:type o:Settlement.
?v1 rdfs:label "Djanet". 
?v2 rdf:type o:Airport. 
?v2 o:city ?v1. 
?v2 o:iataLocationIdentifier ?v3.
?v2 o:location ?v1. 
?v2 p:iata ?v3. 
?v2 foaf:homepage ?v5.
?v2 p:nativename ?v4. 
 \} 

\end{scriptsize}

\end{document}